\documentclass[pre,showpacs, amsmath, amsfonts, amssymb, twocolumn, floatfix]{revtex4}
\usepackage{graphicx}

\newcommand{\vect}[1]{{\mathrm {\mathbf #1}}} 
\newcommand{\real}[1]{{\mathrm Re}\, #1} 
\newcommand{\imag}[1]{{\mathrm Im}\, #1} 
\newcommand{\pderiv}[2]{\frac{\partial #1}{\partial #2}} 
\newcommand{\pdderiv}[2]{\frac{\partial^2 #1}{\partial #2^2}} 

\begin{document}
\title{Complexity of Polarized Spatial Patterns in Large Area Square VCSEL}

\author{I.V. Babushkin}
\affiliation{Max Born Institute for Nonlinear Optics and Short Pulse
Spectroscopy, Max-Born-Str., 2a, D-12489, Berlin, Germany; Fax: +49
30 63921289; e-mail: ibabushkin@mbi-berlin.de}

\author{N.A. Loiko}
\affiliation{Institute of Physics, Academy of Sciences of Belarus,
Scaryna Prospekt 70, 220072 Minsk, BELARUS; Fax: +375-172-393131;
e-mail:nloiko@dragon.bas-net.by}

\author{T. Ackemann}
\affiliation{SUPA and Department of Physics University of
Strathclyde, John Anderson Building, JA 8.21 107, Rottenrow, Glasgow
G4 ONG, Scotland, UK; Fax: +44-(0)141-552 2891; e-mail:
thorsten.ackemann@strath.ac.uk}

\date{\today}

\begin{abstract}
We consider pattern selection process in a wide aperture VCSEL
near threshold. We show that  for a square geometry of the laser
aperture, the patterns formed at lasing threshold can be very
complicated because of a possible misalignment between directions
of an intrinsic spatial anisotropy of VCSEL and lateral boundaries
of its aperture. The analogy with quantum billiard structures is
established, and fingerprints of wave chaos are found. Influence
of localized inhomogeneous in the pump current is also considered.
\end{abstract}


\maketitle

\section{Introduction}

In the last decades, semiconductor laser devices have played an
increasing role in scientific research and applications. Among them,
vertical cavity surface emitting lasers (VCSELs) can be
distinguished, since these lasers emit normal to the wafer surface
in the direction of epitaxial growth. One of the features of VCSELs
design is a possibility to mount extremely large (up to hundreds
$\mu m$ in diameter) aperture with high level of spatial
homogeneity. Spatial mode structures in such wide- aperture VCSELs
have been a subject of many researches
\cite{hegarty99,scheuer99,ackemann00,huang02,chang90,lihua94,
hoersch96,grabherr98,degen01,loiko01b,loiko01}.

In general, the mechanism of pattern formation in VCSEL is very
complicated and involves both a complex structure of a VCSEL cavity
(including Bragg reflectors) \cite{babushkin04,loiko01,loiko01b}, as
well as peculiarities of light-matter interaction in active quantum
well semiconductor layers
\cite{burak00,burak00a,degen00,degen00b,ning95,roessler98}. However,
near threshold one can invoke the perturbation theory and obtain
normal forms governing the evolution of the system. It was shown
recently, that due to a slight spatial anisotropy of a VCSEL cavity
only a few spatial modes come into play at threshold and the shape
of these modes can be analyzed by the linear approximation of the
normal form \cite{loiko01,loiko01b,babushkin04} (in contrast to
spatially isotropic systems, where the whole degenerate family of
modes have the same critical growth rate at threshold, and the
selection process requires consideration of nonlinear competition
even at threshold \cite{lega95,arecchi99}). The investigations of
the problem using this point of view was started in
\cite{babushkin04} for VCSELs with circular aperture. In that work,
transition from the 'flower-like' modes dictated by circular
boundaries to 'stripe-like' ones, which are required by spatial
anisotropy of a VCSEL, was described.

In \cite{babushkin04}, the description was restricted to the light
linear polarized in direction coinciding with one of axes of the
intrinsic laser anisotropy. In the present article, we extend this
approach to the more general "vectorial" case when the laser
anisotropy is not so strong and both orthogonally polarized
components of the field must be taken into account. This extension
allows us to investigate the interaction of two polarization degrees
of freedom, that is important for the consideration of competition
of different mechanisms affecting pattern formation.

The current research is motivated by the recent experimental
investigations \cite{babushkin04, ruhteberg05, ruhteberg06}, where
it is shown that many aspects of spatial structures in VCSEL such as
their transverse shape or frequency-versus-lengthscale dependence
can be well described in a linear approximation, which validates the
'linearized normal form' approach, used in this article.

We show here that this makes the structures at the laser threshold
very complicated even in the case of a simple square aperture and
perfectly manufactured VCSEL cavity. The similar complexity was
observed in the resent experimental investigations
\cite{huang02,chen02,chen03}, where the scarred structures in the
square broad-area VCSEL were obtained and treated as coherent states
of quantum billiard. So, it is possible to speak in this context
about 'quantum chaos' in a wide aperture VCSEL.

The 'quantum chaos' (or 'wave chaos') approach, including the
investigation of level (or eigenfrequency) distribution in a
domain with 'integrable' or 'nonintegrable' boundary conditions,
was applied for investigations of microwave billiards
\cite{stoeckmann99,stoeckmann90,graef92}, as well as semiconductor
microdisk lasers \cite{noeckel97,harayama03}. Very recently, the
wave chaos was observed in a VCSEL with deformed circular cavity
geometry \cite{gensty05}. Though the laser emission far from
threshold is strongly affected by nonlinearities, the optical
spectra obey the same relations as a quantum billiard, governed by
a linear Schr\"odinger equation.

We develop this approach further by considering the wide-aperture
VCSEL near threshold as a kind of quantum billiard where the role
of Hamilton operator is played by the operator of linearized order
parameter equation for VCSEL. We show, that this operator is
considerably more complicated than the transverse Laplace operator
which is often used for consideration of quantum billiard systems.
This leads to appearance of quantum chaos-like features, such as
scarred orbits and nongaussian statistics of the levels even in
the simple square geometry without a deformation of boundary
conditions, that usually used for demonstrating quantum chaos
features.

The structure of the article is following: in the section
\ref{sec:eigen} we extend the approach used in \cite{babushkin04}
to take into account both orthogonally polarized components of the
light field. In the section \ref{sec:comp:struct} we consider the
pattern formation in VCSEL with a square geometry and discuss the
influence of the polarization axes rotation against lateral
boundaries of the cavity, as well as of slight spatial
perturbations of the pumping current. In the section
\ref{sec:chaos} we interpret our results in the 'quantum chaos'
framework.

\section{\label{sec:eigen} The vectorial eigenvalue problem}

As has been pointed in the Introduction, the eigenmodes and their
decay (or growth) rates can be directly obtained from a linear
stability analysis of the nonlasing zero solution of the nonlinear
equations governing VCSEL's dynamics
\cite{loiko01,loiko01b,babushkin04}. In comparison with a method
when the field is decomposed into transverse eigenfunctions of the
empty cavity \cite{coldren95,degen01,mulet02}, this approach takes
into account properties of a gain medium and of Bragg mirrors
composing the cavity by the strick way.

For the infinitely large transverse-area devices the resulting
eigenmodes are plane tilted waves (transverse Fourier modes)  with
the decay rates depending on the detuning of the longitudinal
cavity resonance from the gain maximum and on a polarization of
modes \cite{loiko01,loiko01b,roessler03}.  In contrast,  the
corresponding modes for the finite devices may be very complicated
even in the scalar case. For example, for the circular aperture
the modes can range from Bessel-like modes to the modes resembling
more tilted waves \cite{babushkin04}.

The basic nonlinear equations for the vectorial case were obtained
in \cite{loiko01} . The short description of these equations is
given in Appendix \ref{app:basic_system}. The linearization
procedure is a generalization of one introduced in
\cite{babushkin04} to vectorial case, and is described in Appendix
\ref{app:linearization}. As in the scalar case, we linearize a
complex nonlocal nonlinear operator, obtaining the linear but
nonlocal pseudodifferential operator (or speaking more precisely,
certain eigenvalue problem for such operator) governing the field at
threshold. Then, we approximate the operator describing the action
of Bragg reflectors in transverse Fourier space and obtain the
operator containing only partial derivatives up to fourth order. The
total operator obtained after such procedure includes main
peculiarities of pattern formation such as the dependence on
detuning between longitudinal resonance of the cavity  and gain peak
frequency, and on the anisotropy of Bragg reflectors. This operator
 has the following form in the $(x,y)$ space:
 \begin{equation}
   \hat{ O} = \sum_{i,j=0}^{4} a_{ij} \frac{\partial^{i+j}}{\partial x^i
   y^j} + g_{00} \delta \mu(x,y) + i l \delta n(x,y).
 \label{lin:op:approx}
 \end{equation}
 In contrast to the scalar case,
for the vectorial case operator $\hat{O}$ is a matrix, acting on
the vectorial field $\vect E = (E_x,E_y)$. Therefore, the
coefficients $a_{ij}$ are also $2 \times 2$ matrices. We also
consider inhomogeneities in the pumping current $\delta \mu$ and
in the index $\delta n$, which play the role of disturbances. In
the following, we mainly restrict ourself to only current
inhomogeneities, because they are easier to be controlled.

Eigenfunctions $\vect E_g$ and eigenvalues $\lambda_g$ of this
operator are obtained by solving an eigenvalue problem:
 \begin{equation}
   \hat{O} \vect E_g(x,y) - \lambda_g \vect E_g(x,y) =
   0.
 \label{lin:eigen}
 \end{equation}
The eigenfunction corresponding to the eigenvalue with the largest
real part determines a spatial field distribution with maximal
growth rate at threshold appearing after onset of generation.
Aperture of the device has been simulated by zero field conditions
on the boundaries. Besides, because the operator $\hat O$ is of
fourth order, one should introduce an additional boundary
condition which contains spatial derivatives of the field (see
Appendix \ref{app:linearization}) to obtain the well posed
eigenvalue problem. This second boundary condition can be chosen
by different ways, but all them leads to approximately the same
result \cite{babushkin04}. Therefore, we select the second
boundary condition appropriately, to simplify the solution (see
Appendix \ref{app:implementation}).

The eigenvalue problem (\ref{lin:eigen}) is a generalization of
more conventional one with transverse Laplace operator
(\ref{eq:shredinger:eq:eigenproblem}). The important difference
however,  is that the operator (\ref{lin:op:approx}) is
anisotropic, with a principal  directions defined by polarization
anisotropy of VCSEL cavity. It should be noted that $x$-component
of the field $E_x$ and $y$-component of the field $E_y$ enters to
Eq.~(\ref{lin:eigen}) as just two components of a single vectorial
eigenfunction $\vect E_g$. In the following, we suppose that the
$x$ component of the field is always strongest one due to the
intrinsic polarization anisotropy. However, the orthogonal
component can not vanish because of the mixing of both components
during propagation in the cavity.

\section{\label{sec:comp:struct} Complex spatial structures in square-shaped VCSEL}

\subsection{\label{sec:hom:square} Influence of alignment of
boundaries and ansotropy direction}

To elucidate the mechanism governing the pattern selection, we first
consider the case of homogeneous pumping and index profile ($\delta
\mu = 0, \, \delta n = 0$). The presence of the aperture is modeled
by zero boundary conditions. In this subsection we show, that the
resulting spatio-temporal distribution strongly depends on the
alignment of the boundary conditions and the main anisitropy axes of
the device.

\begin{figure}[htbp!]
\includegraphics[width=75mm]{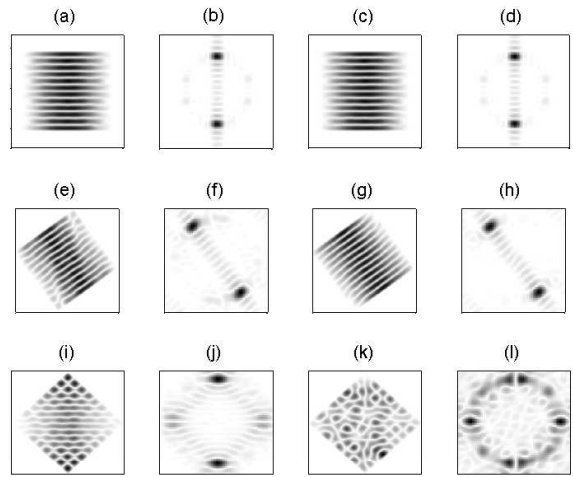}
\caption{ \label{fig-rotate-boundaries}
The structures at threshold obtained as the eigenvectors of
$\hat{O}_{(x,y)}$ with the largest growth rate and different angles
of rotation of boundary against anisotropy direction $\alpha$ (value
of $abs(e)=\sqrt{I}$ is plotted). (a)-(d) --- anisotropy and
boundaries are aligned ($\alpha=0$); (e)-(h) ---
 the misalignment present ($\alpha=\pi/5$); (i)-(l) --- the misalignment
 has its largest possible value ($\alpha=\pi/4$). (a),(e),(i)
--- $x$- component of the field polarization, (b),(f),(j) --- its
transverse Fourier transform; (c),(g),(k) --- $y$- component of
the field polarization, (d),(h),(l) --- its transverse Fourier
transform.  The other relevant parameters are: $l=40 \mu m$
$\gamma_a = 0.01$, $\gamma_p = 0$, $\delta=30 nm$.
} \end{figure}

When boundaries and anisotropy directions are completely aligned,
the $x$-component of the first mode at threshold is a standing
wave with stripes parallel to the direction of its polarization
(see Fig.~\ref{fig-rotate-boundaries}(a), (b)). Though the
y-component is not zero due to the polarization mixing effect by
Bragg reflectors, the coupling between $x$ and $y$ polarizations
is very week since for this case  the Fourier transform of
homogeneous part of $\hat{O}$,
 \begin{equation}
   O_{(k)} = \sum_{i,j=0}^{4} (-i)^{i+j} a_{ij} k_x^i k_y^j.
 \label{lin:op:approx:k}
 \end{equation}
which is $(k_x,k_y)$-dependent 2x2 matrix, have nearly zero
diagonal elements on the line $k_y=0$ and $k_x=0$. Thus,
$y$-component is weaker in 100 times approximately for taken
parameters.

If the boundaries are rotated with respect to the polarization
anisotropy direction, the stripes remain for a small rotation
angle $\alpha$, and their direction coincides with the boundaries
rather then with anisotropy direction, i.e. they are rotated with
the boundaries (see Fig.~\ref{fig-rotate-boundaries}(e)-(h)).
Structures of both polarized components are similar and their
intensity become comparable.

However, as $\alpha$ reaches some critical value, the corresponding
structure becomes more complicated. Thus, for the parameter of
Fig.~\ref{fig-rotate-boundaries}(i)-(l), the spatial distribution of
the $x$-polarized component of the field resembles stripes in the
middle of the aperture, whereas near boundaries the pattern is more
'square-like' one. At the same time, the structure of $y$-polarized
component is more complicated and less regular. The critical angle
for which stripes are giving place to complicated structures
decreases with the size of the device, as well as with a value of
anisotropy $\gamma_a$. However, stripes take place for relatively
large size (composing of several tens of oscillation of the
intensity).

It should be noted that the structures presented in
Fig.~\ref{fig-rotate-boundaries}(i)-(l) are the combination of many
transverse Fourier harmonics of different directions (see
Fig.~\ref{fig-rotate-boundaries}(j),(l)) in contrast to two
travelling waves composing  stripes for small $\alpha$
(Fig.~\ref{fig-rotate-boundaries}(b),(f)).We show in the next
section by considering the corresponding eigenvalue statistics, that
such structures demonstrate some features which make them close to
quantum chaos wavefunctions .

\subsection{\label{sec:inhomogen} Influence of inhomogeneities}

In this subsection, we consider an influence of pump
inhomogeneities of different shape, amplitude and localization. As
it was found earlier for devices with circular aperture
\cite{babushkin03}, any slight inhomogeneities can change the
sequence of a few first eigenfunctions in accordance with the
order of their decay rates, preserving the shape of modes and
their frequencies. In this connection, it is useful to investigate
the subsequent eigenmodes of the linear eigenvalue problem
(\ref{lin:eigen}).

\begin{figure}[htbp!]
\includegraphics[width=75mm]{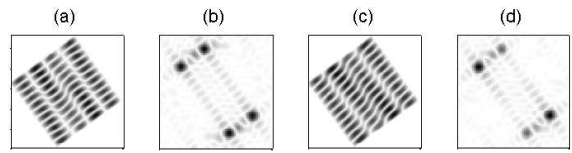}
\caption{ \label{fig-rot-second-mode}
The second mode of homogeneous cavity with parameters as in
Fig.\ref{fig-rotate-boundaries} (e)-(h).  It can be made first by
adding small inhomogeneity in current $\delta \mu$, which is
localized near right boundary of the device.
} \end{figure}

The second eigenmode for the parameters of
Fig.~\ref{fig-rotate-boundaries}(e)-(h) is shown in
Fig.\ref{fig-rot-second-mode}. The Fourier spectrum of this mode
for both polarization components consists of four points: two
pairs of two closed points. These pairs are situated approximately
at the same places that two spots in
Fig.\ref{fig-rotate-boundaries} (f), (h) and rotated by some angle
against polarization direction. So, the main difference between
two first modes is the splitting of the Fourier harmonics leading
to the strong modulation of stripes in the near field
(Fig.\ref{fig-rot-second-mode} (b), (d)). The second eigenmode can
be made leading one (one with the highest growth rate) by adding
small inhomogeneity to the current $\delta \mu = 0.05
\delta(x,y)$, where $\delta(x,y)$ is a probe function which is
nonzero in the vicinity of the right border and zero everywhere
else.

\begin{figure}[htbp!]
\includegraphics[width=75mm]{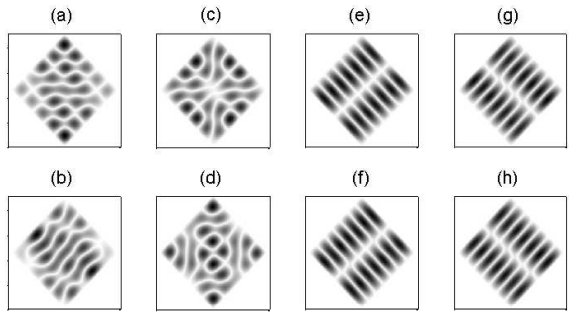}
\caption{ \label{fig-first-egenmodes}
The first (a),(b); second (c,d); 5th and 6th eigenmodes of the
device with the same parameters as in
Fig.~\ref{fig-rotate-boundaries} but with smaller detuning
($\delta=10 nm$), and angle of rotation of boundaries against
anisotropy direction is $\alpha=\pi/4$. (a),(c),(e),(g) - $x$
component of the field, (b),(d),(f),(h) - $y$ component of the
field.
} \end{figure}

If we decrease the detuning $\delta$ between the gain maxima and the
cavity resonance, the gain-loss dispersion mechanism of Fourier
harmonics  selection becomes weaker \cite{loiko01,loiko01b} and, as
result, eigenmodes are more strongly determined by boundary
conditions. The first mode for the large angle near $\pi/4$ does not
have clear stripelike structures in the center of device
(Fig.~\ref{fig-first-egenmodes} (a),(b)), and the second mode has
the structure even less similar to stripes than previous one
(Fig.~\ref{fig-first-egenmodes} (c),(d)). However, stripelike
structures do not disappear at all, as one can see in
Fig.~\ref{fig-first-egenmodes} (e),(f) for the next eigenmode.
Moreover, the patterns with orthogonal direction of stripes can be
easily exited (Fig.~\ref{fig-first-egenmodes} (g),(h)). This could
be explained by the fact that these two families of stripes
 are degenerate when $\alpha = 0$, and start to
deviate from each other with increasing $\alpha$.

\begin{figure}[htbp!]
\includegraphics[width=75mm]{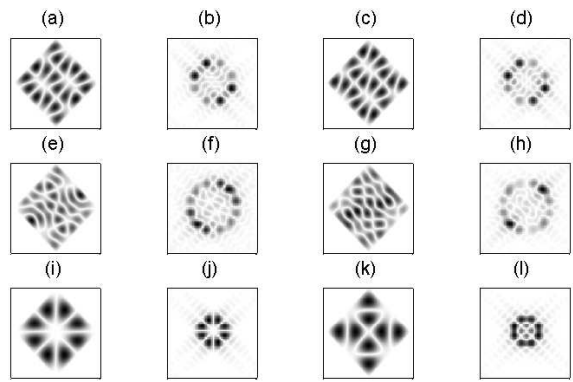}
\caption{ \label{fig-subseq-egenmodes}
The subsequent eigenmodes for the same parameters as in
Fig.~\ref{fig-first-egenmodes}. (a)-(d) - 7th (e)-(h) - 21th,
(i)-(l) - 26th eigenmode. (a),(e),(i) --- $x$- component of the
field polarization, (b),(f),(j) --- its transverse Fourier
transform; (c),(g),(k) --- $y$- component of the field polarization,
(d),(h),(l) --- its transverse Fourier transform.
} \end{figure}

Some subsequent eigenmodes with larger decay rate for the same
parameters are presented in Fig.~\ref{fig-subseq-egenmodes}. It is
evident that for small enough (but nonzero) anisotropy the modes
can differ from each other in their symmetry properties
essentially, as in the case of circular aperture
\cite{babushkin04}. One can find nearly unordered structures
(Fig.~\ref{fig-subseq-egenmodes} (e)-(h)) and the structures which
can be considered as 'scared' eigenmodes of quantum billiard
(Fig.~\ref{fig-subseq-egenmodes} (a)-(d), (i)-(l)).

It is difficult to change the order of subsequent modes
(especially to push one of them to the first position with highest
grough rate) by introducing a small inhomogeneity. Since stronger
inhomogeneities needed for that mode structures can be changed
also.

It should be also noted that whereas patterns of the
orthogonally-polarized component (weaker component) of the laser
field are usually of the same shape, one can find some exceptions
(compare Fig.~\ref{fig-subseq-egenmodes}(i) and
Fig.~\ref{fig-subseq-egenmodes}(k)). However, even in this case the
transverse Fourier images of the field components look quite
similar. This is connected to the fact, that structures in both
polarizations belong to the same eigenmode and coupled to each other
in the absence of strong inhomogeneities through boundary conditions
(they both and some combination of their derivatives must vanish at
the boundary), and through the Bragg reflectors. The exception is
only when this connection is very weak
(Fig.~\ref{fig-rotate-boundaries} (a)-(d)).

\begin{figure}[htbp!]
\includegraphics[width=75mm]{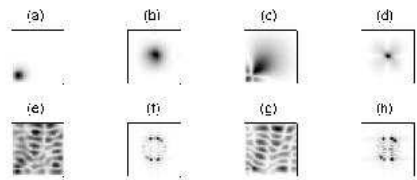}
\caption{ \label{fig-peak-in-aperture}
The first two eigenmodes for the device with the same parameters
as in Fig.~\ref{fig-rotate-boundaries} (a)-(d), but $\delta=8nm$,
and peak of $\delta \mu$ in the aperture. (a),(e) --- $x$-
component of the field polarization, (b),(f) --- its transverse
Fourier transform; (c),(g) --- $y$- component of the field
polarization, (d),(h) --- its transverse Fourier transform.
} \end{figure}

Up to now we have investigated the eigenmodes of homogeneous
devise, having in mind that the grough rate of a mode following
the first one can become the largest one when  a small
perturbation of the current $\delta \mu$ is added, which does not
sufficiently change the shape of mode.

Increasing the intensity of inhomogeneities leads to changes in
the shape of eigenmodes. As an example, we consider here strong
inhomogeneities by introducing a peak or a hole into the laser
aperture, that is quite natural for experiments.

For the peak in current profile, the first eigenmode is defined by
this spot only, as if the rest of aperture  is empty
(Fig.~\ref{fig-peak-in-aperture} (a)-(d)). But, the subsequent
modes fill all the cavity as there is no peak in the aperture at
all (Fig.~\ref{fig-peak-in-aperture} (e)-(h)). However, they are
more disordered then in the absence of inhomogeneity. The
corresponding eigenvalues are also more separated as it is
described in the next section.

\begin{figure}[htbp!]
\includegraphics[width=75mm]{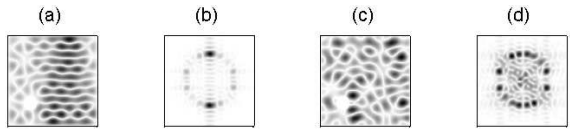}
\caption{ \label{fig-hole-in-aperture}
The first eigenmode for the same parameters as in
Fig.~\ref{fig-rotate-boundaries} (a)-(d), but the boundary
conditions (\ref{lin:op:bound:cond:def}) are defined not on the
boundaries of a square, but also on a circle inside the square
(which creates the hole in the aperture). The hole is of $1/8$ of
the size of the laser and placed in the position $(1/4,1/4)$ in
the whole aperture.
(a) --- $x$- component of the field
polarization, (b)
--- its transverse Fourier transform; (c) --- $y$-
component of the field polarization, (d) --- its transverse
Fourier transform.
} \end{figure}

On the other hand, if there is a strong hole in the current profile
(which is for simplicity modeled by introducing zero boundary
conditions on the circle surrounding the hole), the effect is not so
visible, i.e. there is not a separation on the "hole" mode and the
rest ones. Moveover, the shape of eigenmodes for $x$ polarization is
not changed so dramatically far from the hole
(Fig.~\ref{fig-hole-in-aperture}), although the structure of weak
component is strongly disordered comparing to the nonperturbed case
(Fig.~\ref{fig-rotate-boundaries}(c)). However, the subsequent
eigenmodes are more irregular that the first one even for the
x-component. The problem becomes close to chaotic one
 that could be considered in the
framework of wave chaos.

\section{\label{sec:chaos} Signatures of quantum chaos}

Generally speaking, the term 'quantum chaos' or 'wave chaos' is
usually attributed to investigation of quantum systems, which
posses chaotic features in classical limit \cite{stoeckmann99}. In
contrast to classical systems, they obey linear evolution equation
(Schr\"odinger equation). So, the object of investigation is
usually the set of eigenvalues and eigenfunctions (stationary
states) of some linear operator, describing the quantum system
under consideration (usually Hamilton operator). At that, the very
important role is played by eigenvalue distribution, which shows
fingerprints of chaos in quantum systems when the corresponding
classical systems are chaotic.

Among the quantum systems, much attention is devoted to the
billiard systems \cite{stoeckmann99,cvitanovic04}. The quantum
counterpart of classical billiard is described by a wavefunction
$\psi(x,y,t)$, which obeys Schr\"odinger equation:
\begin{equation}\label{eq:shredinger:eq:billiard}
    i \hbar \pderiv{\psi}{t} = \Delta \psi,
\end{equation}
with boundary condition
\begin{equation}\label{eq:shredinger:eq:billiard:boundary:cond}
    \psi = \left.0\right|_{\partial S},
\end{equation}
on the boundary of the domain $S$.

Stationary states of this problem are eigenfunctions
 of the transverse Laplace operator $\Delta = \pdderiv{ }{x} + \pdderiv{
 }{y}$:
\begin{equation}\label{eq:shredinger:eq:eigenproblem}
    i\Delta \psi = \lambda \psi.
\end{equation}

 Depending on the shape of domain $S$, the corresponding
 classical problem can be either chaotic or regular. In the
 limit case of integrable system (for example, when the region has a rectangular shape),
 the corresponding eigenvalue problem gives set of eigenvalues, obeying
 Poisson statistics of an eigenvalue spacing $s$:
\begin{equation}\label{eq:puasson:statistic}
    p(s) = exp(-s),
\end{equation}
where $p(s)$ is a probability for corresponding value of
eigenfrequency distribution $s_i= ~ \imag \lambda_i - \imag
\lambda_{i+1}$.

For the opposite limit case of completely chaotic system, the
corresponding distribution is the Wigner one:
\begin{equation}\label{eq:wigner:statistic}
    p(s) = \frac{1}{2}\pi s exp(-\frac{1}{4} \pi s^2).
\end{equation}

For intermediate situations one can define other statistic
families limited by Eq.~(\ref{eq:puasson:statistic}) and
Eq.~(\ref{eq:wigner:statistic})
\cite{stoeckmann99,izrailev88,brody73}. One of fingerprints of
quantum chaos in these intermediate cases, as well as in the Eq.
(\ref{eq:wigner:statistic}), that the maximum of $p(s)$ is reached
for $s \ne 0$. This defines so called level repulsion phenomena
for chaotic systems.

The eigenfunctions of Hamilton operator also demonstrate
fingerprints of chaos. In particular, one can observe so called
'scarred' patterns
\cite{heller84,bogomolny88,berry89,chen02,chen03}, which are
localized near (unstable) periodic trajectories of corresponding
classical systems. Among the scared patterns, there are other
eigenfunctions, which are strongly irregular and fit all the area.

It should be noted, that besides the quantum systems, the above
mentioned framework is of the considerable interest for
macro-systems, which are described at some level of approximation by
Eq.~(\ref{eq:shredinger:eq:eigenproblem}) or
Eq.~(\ref{eq:shredinger:eq:billiard}). Among them are microwave
billiards \cite{stoeckmann99,stoeckmann90,graef92}, microdisk lasers
\cite{noeckel97,harayama03} and VCSELs
\cite{gensty05,chen02,chen03}.

The eigenvalue statistics and other fingerprints of quantum chaos
can give a criterion of a complexity of patterns near threshold.
Indeed, if a system possess the Poisson statistics, the
corresponding eigenfunctions are regular and stable against small
perturbations, whereas for the Wigner one the eigenfunctions are
much more irregular.

\begin{figure}[htbp!]
\includegraphics[width=75mm]{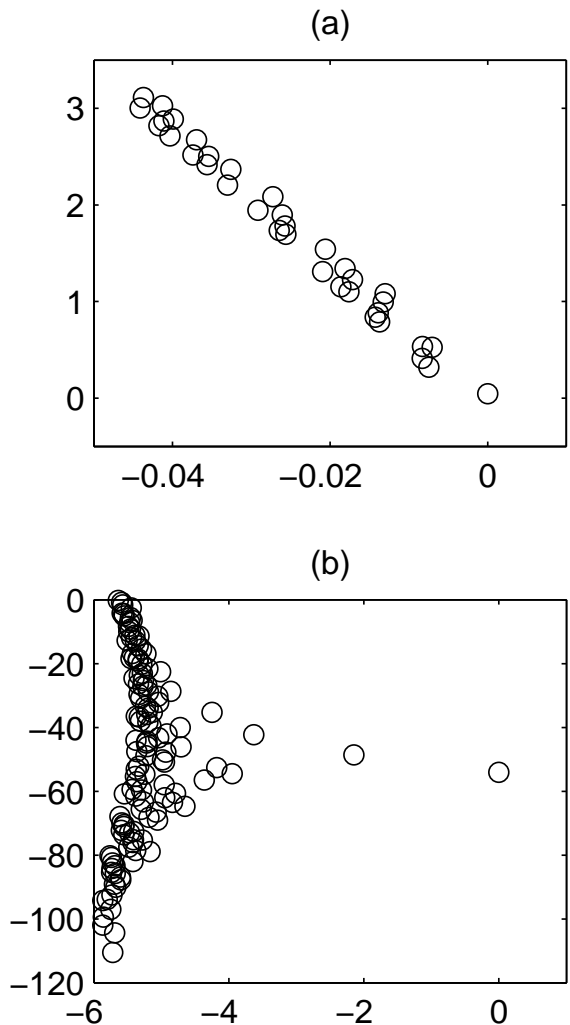}
\caption{ \label{fig-reim-parts}
(a) --- Imaginary versus real part of few tens eigenvalues
$\lambda_i$ of operator $\hat O$ having the largest value of
$\real \lambda_i$ for VCSEL with parameters as in
Fig.~\ref{fig-rotate-boundaries} (e)-(h). (b) --- Imaginary versus
real part of about hundred eigenvalues $\lambda_i$ of $\hat O$ for
parameters close to ones used for Fig.~\ref{fig-peak-in-aperture}.
} \end{figure}

In the present work, we describe the pattern formation in VCSEL by
linear operator $\hat O$ (\ref{lin:op:approx}) obeying
Eq.~(\ref{lin:eigen}) This operator is considerably more
complicated then the transverse Laplace operator from
Eq.~(\ref{eq:shredinger:eq:eigenproblem}). First of all, the
eigenvalues of the operator $\hat O$ are complex. Their real parts
describe the decay rate of the corresponding eigenfunctions,
whereas the complex parts are their oscillation frequencies
determining an energy level distribution. Examples of the
eigenvalues distribution are presented in
Fig.~\ref{fig-reim-parts}. For homogeneous device, all eigenvalues
are concentrated  along one line and in average their decay rates
are increased with the frequencies (Fig.~\ref{fig-reim-parts}
(a)). The closeness of decay rates of the adjacent modes allows to
change their order by a small perturbation. It is worth noting
that for the taken parameters the eigenvalues are grouped in
clusters and become deviate with decreasing of the angle $\alpha$.
In the case of a local peak in the pump current the eigenvalues
distribution confirms the separation of the first and the rest
modes (Fig.~\ref{fig-reim-parts} (b)).


\begin{figure}[htbp!]
\includegraphics[width=75mm]{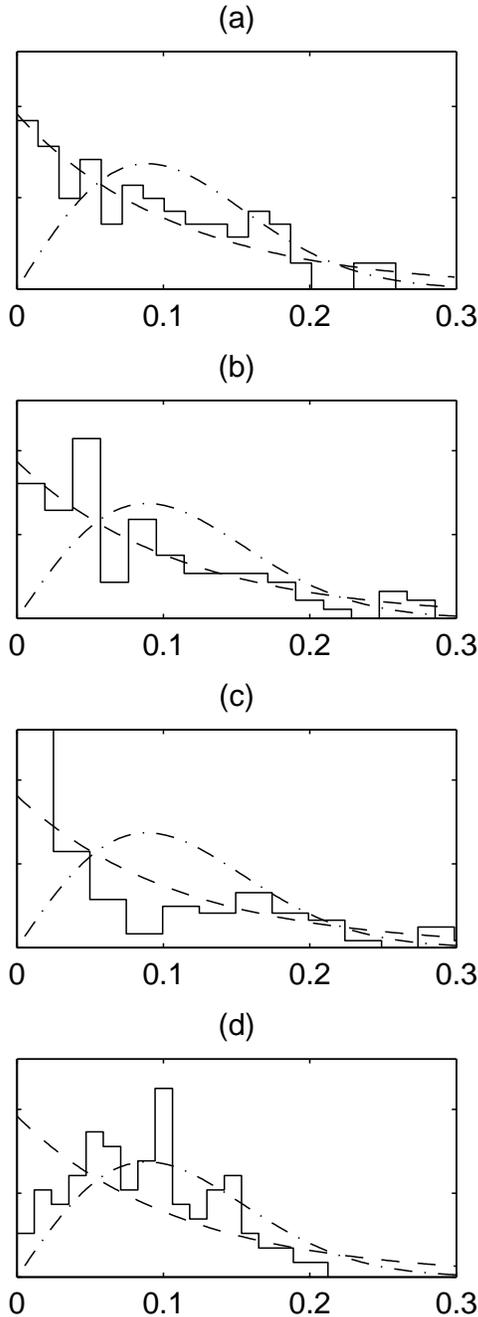}
\caption{ \label{fig-statistics}
The statistics of approximately 100 first eigenmodes (with largest
value of $\real \lambda$) of VCSEL with parameters and boundaries
as in Fig.~\ref{fig-rotate-boundaries}  and
Fig.~\ref{fig-hole-in-aperture},  shown as stairstep graph, bold
line. For a comparison Poisson statistics
(Eq.~\ref{eq:puasson:statistic}) Wigner statistics
(Eq.~\ref{eq:wigner:statistic}) is shown by dashed and dot-dashed
lines, correspondingly. (a) corresponds to (a)-(d) in
Fig.~\ref{fig-rotate-boundaries}, i.e. $\alpha=0$, (b) corresponds
to (e)-(h) in Fig.~\ref{fig-rotate-boundaries}, i.e.
$\alpha=\pi/5$, (c) corresponds to (i)-(l) in
Fig.~\ref{fig-rotate-boundaries}, i.e. $\alpha=\pi/4$, (d)
corresponds to complicated region of
Fig.~\ref{fig-hole-in-aperture}.
} \end{figure}

On the other hand, the operator (\ref{lin:op:approx}) is
anisotropic, with a preferential directions defined by anisotropy
of VCSEL cavity. As one can see from
(\ref{lin:op:approx}),(\ref{lin:eigen}), if the boundaries are
defined parallel to $x$ and $y$ axes, one can separate variables
in the eigenvalue problem and therefore it becomes integrable.
Hence, as one can see in Fig.~\ref{fig-statistics}(a), when the
boundaries and anisotropy direction coincides, the eigenvalues
obey more or less Poisson statistics.

When these two directions (anisotropy and boundaries for square
shape of the aperture) are not aligned, the variable separation is
not possible anymore (as would be for the second order differential
operator). As a result, for $\alpha \ne 0$
(Fig.~\ref{fig-statistics}(b)) one can clearly see that the maximum
probability of the level spacing distribution is shifted to nonzero
$s$, possessing a fingerprint of wave chaos. However, the system can
not be considered as purely 'quantum chaotic' because the statistic
is of intermediate type, neither purely Poisson nor Wigner one. It
should be noted, that for isotropic system obeying
Eq.~(\ref{eq:shredinger:eq:eigenproblem}) the eigenvalue spacing
statistics remain Poisson for any values of $\alpha$. In our case,
 the eigenvalue spacing has the maximum
value again at zero for the angle $\alpha=\pi/4$, that can be
explained by possible degeneracy of the operator $\hat O$
(Fig.~\ref{fig-statistics}(c)).

As a comparison the eigenlevel statistics is shown for boundary
conditions of Fig.~\ref{fig-hole-in-aperture}. It is clearly seen
that for this case the statistics for spacing of imaginary parts
of eigenvalues can be considered as Wigner one, which is quite
expectable, because the same happens with the eigenvalues of the
transverse Laplace operator. On the other hand, for the large peak
in the aperture the eigenvalues do not obey the Wigner statistics
anymore. For this case, the leading eigenvalues (eigenfunctions
for which are shown in Fig.~\ref{fig-peak-in-aperture}(a)-(d) and
can be described as eigenvalues of a disturbance only) are
strongly separated from others (which are resembling
Fig.~\ref{fig-peak-in-aperture}(e)-(h)), which obviously breaks
the Wigner distribution.

The other signatures of complexity of spatial structures can be
found in fingerprints of quantum chaos demonstrated by the shape of
eigenfunctions. As one can see, the neighboring eigenmodes for
misaligned boundaries and anisotropy directions ($\alpha \ne 0$) can
differ sufficiently (Fig.~\ref{fig-first-egenmodes},
Fig.~\ref{fig-subseq-egenmodes}), including very irregular ones
(Fig.~\ref{fig-subseq-egenmodes}(e)-(g)), which is a common property
of systems possessing quantum chaos. On the other hand, beside
complicated unordered structures, quantum chaos is characterized by
'scared' eigenfunctions, which are located near unstable periodic
trajectories of corresponding classical system. The example of
eigenfunctions of operator $\hat O$, resembling that kind of
structures (Fig.~\ref{fig-subseq-egenmodes})(a)-(d),(i)-(l).
Experimentally these structures were observed recently in VCSEL
\cite{chen02,chen03}.

\section{\label{sec:conclusion} Discussion and Conclusion}

We have considered the pattern formation in wide aperture VCSEL near
lasing threshold. Because of the Bragg reflectors, the presence of
polarization anisotropy creates a spatial anisotropy. This allows to
investigate the structure of pattern near threshold by linearized
order parameter equations, since even in a linear approximation the
mode with the smallest decay rate is nondegenerate (in contrast to
spatially isotropic systems, where the whole family of modes has the
same growth rate at threshold, and nonlinear competition is of
strong importance for the selection). Since our studies concentrated
on a VCSEL with square aperture, the competition of boundary
direction (which can be defined as a vector, parallel to one of the
sides of the square), and spatial anisotropy is important mechanism
for pattern selection.

To measure the complexness of a pattern structure we use criteria
from the topic of quantum chaos, where the main object of research
is also an eigenvalue problem of some operator. One of
fingerprints of complicated behavior is an eigenvalue spacing
statistics.

In this work we have considered a statistics of imaginary parts of
eigenvalues, describing frequencies of corresponding
eigenfunctions. We have shown that in the case of aligned
directions of boundaries and spatial (or polarization) anisotropy,
the above mentioned statistics is close to Poisson one. However,
when the angle of rotation of spatial anisotropy against
boundaries is not zero, the statistics is not Poisson anymore. In
the later case, the statistics shows a signature of level spacing
repulsion, i.e. maximum of eigenspacing distribution is not zero.

The other fingenprints of quantum chaos can be found in the
spatial shape of eigenfunctions, which can be very irregular, as
well as 'scared-like' patterns.

To compare these results with experiments, we also have
investigated the influence of the inhomogeneities. The considered
system is stable in the sense, that small perturbations only
change the modes order preserving their shape, and not lead to
dramatic change of the whole picture. In this sense, the
consideration of subsequent modes (having higher decay rate) are
also important, because they can be considered as leading ones of
the device perturbed by some (may be unknown) inhomogeneity.

With increasing the inhomogeneity, the shape of eigenmodes also
changes, as well as all the characteristics like the level spacing
distribution. One can, for example, model the hole in the current
by imposing zero boundary conditions somewhere on a circle located
inside the squire aperture. In this case, the statistics evidently
becomes Wigner, because the corresponding billiard is scattering
one. On the other hand, large positive peak in the current leads
to separation of eigenvalues and eigenfunctions into two classes,
first of them belonging to 'peak family', with eigenfunctions
localized near the peak, whereas the second family is distributed
across the whole cavity aperture.

\acknowledgments

This work was financially supported by the Deutsche
Forschungsgemeinschaft for equipment and by travel grants. We are
grateful to Malte Schutz-Ruhtenberg for useful discussions.

\appendix

\section{\label{app:basic_system} Basic System}

The system under consideration was initially obtained in
\cite{loiko01}. The equations for the field $\vect E = (E_x,E_y)$
take into account propagation inside the complex VCSEL cavity. The
material equations are based on spin-flip model
\cite{sanmiguel99}, describing four level system with populations
differences of relevant transitions determined by their sum $N$
and difference $n$:

\begin{widetext}
\begin{eqnarray}
\label{eq:app:e}
  \frac{{d\mathbf{E}(x_t ,t)}}
{{dt}} & = &   - \kappa \hat M\mathbf{E} + i\hat \Omega \mathbf{E} - \hat \Gamma \mathbf{E} - i\kappa \alpha \mathbf{E} + \kappa (1 + i\alpha )\hat G\hat L_\Omega  (\hat A\mathbf{E}), \hfill \\
  \label{eq:app:dN} \frac{{dN}}
{{dt}} & = &  - N + \mu - \operatorname{Im} [(i - \alpha )\mathbf{E}^* \hat L_\Omega  (\hat A\mathbf{E})], \hfill \\
 \label{eq:app:dn} \frac{{dn}}
{{dt}} & = &  - \gamma _s d - \operatorname{Re} [(i - \alpha )\mathbf{E}^* \hat L_\Omega  (\hat A\mathbf{E})], \hfill \\
\end{eqnarray}
\end{widetext}
where
$$
\mathbf{{\rm A}} = \left( {\begin{array}{*{20}c}
   {N } & {in}  \\
   { - in} & {N }  \\
 \end{array} } \right),
$$
and the operators $\hat M$ and  $\hat G$ describes the modal
losses and gain, correspondingly, $\hat \Omega$ describes the
influence of the diffraction in the complex resonator of VCSEL.
These operators take into account the influence of reflection from
Bragg reflectors and are quite complicated. More details can be
found in \cite{loiko01,babushkin04}.

The operator
\[
\hat L_\Omega   = 1/\left( {1 + \left( {\frac{{\delta  - \hat
\Omega }} {\gamma }} \right)^2 } \right)
\]
describes the Lorentz shape of gain contour with a detuning
$\delta$ between the gain maximum and cavity resonance frequency.
$\kappa$ is the field decay rate in the VCSEL's cavity.

The operator $\hat \Gamma$ describes the inner anisotropy of
VCSEL's cavity:

\[
\mathbf{\hat \Gamma } = \left( {\begin{array}{*{20}c}
   {\exp (\gamma _a  + i\gamma _p )} & 0  \\
   0 & {\exp ( - (\gamma _a  + i\gamma _p ))}  \\

 \end{array} } \right)
\]
where $\gamma_a$ and $\gamma_p$ is an amplitude and phase
anisotropy, respectively.

\section{\label{app:linearization} Vectorial Eigenvalue Problem Derivation}

To derive the linear evolution equation in the form,
 \begin{equation}
   \pderiv{E}{t}  = \hat{O} \vect E_g(x,y),
 \label{eq:app:lin:eigen}
 \end{equation}
(which is then reduced to Eq.~\ref{lin:eigen}) using the basic
equations (\ref{eq:app:e}-\ref{eq:app:dn}), we take into account
that at the laser threshold two branches of the steady state
solutions cross each other: the zero solution ($\vect E=0$), and
the nonzero lasing one. Because of the lasing solution at the
cross-section point is characterized by $\vect E=0$ and $N=\mu$
the further analysis is drastically simplified giving the
following diagonal operator of the linearized problem:
 \begin{equation}
 \hat{ T} = \left(
 \begin{array}{cc}  \hat{\vect O}_{11} & 0
 \\ 0 &  \hat{\vect O}_{22}
 \end{array} \right).
 \label{lin:op:2D:gen:form}
 \end{equation}
where each $\hat{ O}_{ii}$ is in turn the matrix operator:

 \begin{equation}
 \hat{ O}_{ii} = \left(
 \begin{array}{cc}  \hat O^{(ii)}_{11} & \hat O^{(ii)}_{12}
 \\ \hat O^{(ii)}_{21} &  \hat O^{(ii)}_{22}
 \end{array} \right).
 \label{eq:op:def:e_part}
 \end{equation}

The diagonal form of (\ref{lin:op:2D:gen:form}) shows, that the
field $\vect E$ and the variables $N$,$n$ are independent of each
other at the laser threshold, and the spectrum for $\hat{ O}_{22}$
lies entirely in the half-plane $\real \lambda \leq 0$, hence we
can consider only the equation for the field
Eq.~\ref{eq:app:lin:eigen} with $\hat{ O} \equiv \hat{ O}_{11}$.
$\hat{O}$ includes the current density at threshold $N = \mu$.
From (\ref{eq:app:e}),(\ref{lin:op:2D:gen:form}), the explicit
value for the operator $\hat{ O}$ is defined by the following
action on the field $\vect E$:
\begin{widetext}
 \begin{equation}
   \hat{ O} = - \kappa \hat M  + i\hat \Omega  - \hat \Gamma  - i\kappa \alpha  + \kappa (1 + i\alpha )\mu \hat G\hat L_\Omega  \hfill.
 \label{lin:op:expl}
 \end{equation}
 \end{widetext}

However, the expression (\ref{lin:op:expl}) is still nonlocal,
because all the operators in (\ref{lin:op:expl}) are
integrodifferential. The equation can be made local in the manner
of \cite{babushkin04}. The procedure used for a scalar operator in
\cite{babushkin04}, must be applied to each of the operators
$O^{(11)}_{ij}$, which are the scalar components of $\hat{ O}$.
Because the diagonal part of (\ref{eq:op:def:e_part}) plays the
main role, it is approximated up to the fourth order of $\vect
k_\bot$, whereas the operators $O^{(11)}_{ij}$ for $i \ne j$ are
approximated with terms of the second order of $\vect k_\bot$.

In the presence of index inhomogeneities, $\hat \Omega =
\Omega_{homogen} + i l \delta n$ (where $l= \tau \omega/n_0$ is
expressed via the optical frequency $\omega$, round trip time of
the cavity $\tau$´and the mean index $n_{0}$. The pump current
inhomogeneities have to be included in the coefficient of the last
term  $\mu= \mu_{homogen} + \delta \mu$.

\section{\label{app:implementation} Numerical Implementation of Eigenvalue Problem}

To solve the problem, the Matlab PDE toolbox implementation
\cite{pdetoolbox} of Arnoldi method for the system of partial
differential equations of the second order has been used. For
that, the operators of the fourth order $\hat O^{(11)}_{ii}$ have
been represented via multiplication of two operators
$\hat{P}_{i1}$, $\hat{P}_{i2}$ of the second order with constant
coefficients:
 \begin{equation}
   \hat O^{(11)}_{ii} = \hat{P}_{i1} \hat{P}_{i2} - l_{i1} =  \hat{P}_{i2} \hat{P}_{i1} -
   l_{i1}
 \label{lin:op:decompos:inhom:multiplication}
 \end{equation}
where $P_{i1} = \vect \nabla \cdot ( c^{(i)}_{12} \otimes \vect
\nabla)$, $P_{i2} = \vect \nabla \cdot ( c^{(i)}_{21} \otimes
\vect \nabla)$.

Here $ c^{(k)}_{ij}$ are matrices defined by formulas given in
\cite{babushkin04} using the coefficients $a_{ij}$ (to define $
c^{(1)}_{ij}$ and $c^{(2)}_{ij}$ the coefficients
$(a_{ij})_{(11)}$ and $(a_{ij})_{(22)}$ are used,
correspondingly). Then, defining a new vector variable
 \begin{equation}
        \vect E_1 = (\hat{P}_{12} E_x,\hat{P}_{22} E_y)
 \label{lin:op:decompos:inhom:e1:def:1}
 \end{equation}
or
 \begin{equation}
        \vect E_1 = (\hat{P}_{11} E_x,\hat{P}_{21} E_y)
 \label{lin:op:decompos:inhom:e1:def:2}
 \end{equation}
one can present the action of the operator of fourth order $\hat{
O}_{(x,y)}$ as a operator of the second order of the type
 \begin{equation}
   \hat {\rm P} = \vect \nabla \cdot \left( c \otimes \vect \nabla \right) +
   a.
 \label{lin:op:decompos:p}
 \end{equation}
acting on a four-component function $\vect E_{ex} = (\vect E,\vect
E_1)$. Here $a$ is a $4 \times  4$ matrix (instead of $2 \times 2$
for scalar case \cite{babushkin04}), $c$ is a rank four tensor
which can be described by eight $2 \times 2$ matrices $
c^{(k)}_{ij}$, mentioned above after equation
(\ref{lin:op:decompos:inhom:multiplication}). The boundary
conditions have to be also generalized for the function  $\vect
E_{ex}$ as in the work \cite{babushkin04}:
 \begin{equation}
     \vect E_{ex} = 0:   \vect E = 0, \vect E_1=0,
 \label{lin:op:bound:cond:def}
 \end{equation}
where $\vect E_1$ is defined by
(\ref{lin:op:decompos:inhom:e1:def:1}) or
(\ref{lin:op:decompos:inhom:e1:def:2}).

 In the framework of the
original system it means that we introduce a second boundary
conditions on the vector field $\vect E$ in accordance with that
the operator $\hat{O}_{(x,y)}$ is of the fourth order.



\end{document}